\documentclass[
    ,final            
  ]
  {aipproc}

\layoutstyle{8x11single}
\usepackage{amsmath}
\usepackage{graphicx}
\usepackage{epsfig}
\usepackage{rotating}

\begin{document}

\title{The phase diagram in the SU(3) Nambu-Jona-Lasinio model with 't Hooft and eight-quark interactions.}


\author{J. Moreira}{
  address={Centro de F\'{i}sica Computacional, Departamento de F\'{i}sica da Universidade de Coimbra,
3004-516 Coimbra, Portugal}
}

\author{B. Hiller}{
  address={Centro de F\'{i}sica Computacional, Departamento de F\'{i}sica da Universidade de Coimbra,
3004-516 Coimbra, Portugal}
}

\author{A. A. Osipov}{
  address={Centro de F\'{i}sica Computacional, Departamento de F\'{i}sica da Universidade de Coimbra,
3004-516 Coimbra, Portugal}
  ,altaddress={Dzhelepov Laboratory of Nuclear Problems, JINR 141980 Dubna, Russia} 
}

\author{A. H. Blin}{
  address={Centro de F\'{i}sica Computacional, Departamento de F\'{i}sica da Universidade de Coimbra,
3004-516 Coimbra, Portugal}
}

\begin{abstract}
It is shown that the endpoint of the first order transition line which merges into a crossover regime in the phase diagram of the Nambu--Jona-Lasinio model, extended to include the six-quark 't Hooft and eight-quark interaction Lagrangians, is pushed towards vanishing chemical potential and higher temperatures with increasing strength of the OZI-violating eight-quark interactions.  We clarify a connection between the location of the endpoint in the phase diagram and the mechanism of chiral symmetry breaking at the quark level. Constraints on the coupling strengths based on groundstate stability and physical considerations are explained.
\end{abstract}

\maketitle


\section{Introduction}

The determination of the phase diagram of QCD represents an interesting challenge both from the experimental and the theoretical points of view. On the experimental side several new facilities are aiming at an unprecedented look at the properties of hot and/or dense strongly interacting matter and from the theoretical point of view, as standard perturbative techniques fail in the low energy regime of QCD, alternative approaches have to be used. Great progress has been achieved over the last two decades in the theoretical study of the QCD phase diagram both in terms of effective low energy theories and lattice calculations (see for instance reviews \cite{Rajagopal:2000wf,Schafer:2005ff,Stephanov:2004xs,Fukushima:2008pe} and paper \cite{Ratti:2005jh}). 

The Nambu--Jona-Lasinio model \cite{Nambu:1961tp, Nambu:1961fr, Vaks:Larkin}, is regarded as an useful tool for the study of low
energy hadron phenomenology as it shares with QCD the global symmetries and incorporates by construction a mechanism for dynamical chiral symmetry breaking (D$\chi$SB). The extension to include a $2N_f$ 't Hooft determinantal interaction (NJLH) \cite{'tHooft:1976fv, PhysRevD.18.2199.3, Bernard:1987gw, Bernard:1987sg, Reinhardt:1988xu} (where $N_f$ is the number of flavors) breaks the unwanted $U_A(1)$ symmetry exhibited by this model. The study of this model in the light quark sector (u, d and s) has shown a fundamental flaw \cite{Osipov:2005tq} (the absence of a globally stable ground state) which can be remedied with the inclusion of general scalar non derivative $U_L(3)\times U_R(3)$ symmetric eight quark interactions. 

\section{Formalism}

The model Lagrangian that we are going to use, $\mathcal{L}_{eff}$, can be decomposed in the following manner (see \eqref{LagrangianDef}):
\begin{align}
\label{LagrangianDef}
\mathcal{L}_{eff} & = \mathcal{L}_D +\mathcal{L}_{NJL}+\mathcal{L}_{H}+\mathcal{L}_{8q} &\qquad &\nonumber\\
\mathcal{L}_D &=\overline{q}\left(\imath\gamma^\mu \partial_\mu-m\right)q &\qquad  
\mathcal{L}_{8q} & =\mathcal{L}^{(1)}_{8q}+\mathcal{L}^{(2)}_{8q} 			\nonumber\\
\mathcal{L}_{NJL}&=\frac{G}{2}\left[\left(\overline{q}\lambda_a q\right)^2 + \left(\overline{q}\imath \gamma_5 \lambda_a q\right)^2\right] & \qquad
\mathcal{L}^{(1)}_{8q} & =8 g_1\left[\left(\overline{q}_iP_{R} q_{m}\right)\left(\overline{q}_mP_L q_i\right)\right]^2 	\nonumber\\
\mathcal{L}_{H} & = \kappa \left(\mathrm{det}\overline{q}P_Lq +\mathrm{det}\overline{q} P_R q \right)& \qquad
\mathcal{L}^{(2)}_{8q} & =16 g_2\left(\overline{q}_iP_R q_m\right)\left(\overline{q}_mP_L q_j\right)
					 \left(\overline{q}_jP_R q_k\right)\left(\overline{q}_kP_L q_i\right).
\end{align}
The free Dirac term, $\mathcal{L}_D$, contains a current mass matrix, $m$, diagonal in flavor space which explicitly breaks chiral symmetry (we will consider $m_u=m_d\neq m_s$ as such $SU(3)$ flavor symmetry is also broken). The 4-quark NJL interaction, $\mathcal{L}_{NJL}$, if strong enough induces spontaneous chiral symmetry breaking 
 (here $\lambda_0=\sqrt{2/3}$ and the $\lambda_a$, $a=1,\ldots,8$ are Gell-Mann matrices acting in flavor space). The 6-quark 't Hooft term, $\mathcal{L}_{H}$, breaks axial symmetry (here the $P_{L,R}$ are chiral projectors and the determinant is over flavor space). The 8-quark interaction, $\mathcal{L}_{8q}$, can be split in two parts, one of which, $\mathcal{L}^{(1)}_{8q}$, along with the 't Hooft term, induces OZI rule violation.

This additional term ensures ground state stability as long as: $g_1>0$ , $g_1+3g_2>0$ and $g_1>1/G\left(\kappa/16\right)^2$ \cite{Osipov:2005tq}. Through a suitable choice of parameters (a decrease in $G$ with increased $g_1$ with remaining parameters kept fixed) the light scalar and pseudoscalar meson spectra in the vacuum can be left relatively unchanged, apart from a marked decrease in the $\sigma$ meson mass, and close to the one obtained in the NJLH model \cite{Osipov:2006ns}. 
\footnote{We chose to fit the model parameters $G,~\kappa,~g_2,~\Lambda,~m_u(=m_d),~m_s$ to reproduce the following vacuum properties: $M_\pi=0.138\mathrm{GeV}$, $M_K=0.494\mathrm{GeV}$, $M_{\eta^\prime}=0.958\mathrm{GeV}$, $M_{a_0}=0.980\mathrm{GeV}$, $f_{\pi}=0.092\mathrm{GeV}$ and $f_{K}=0.117\mathrm{GeV}$. This sets the stability bounds on the value for the coupling strength $347\mathrm{GeV}^{-8}=g^{(1)}_1<g_1<g^{(2)}_1=20300\mathrm{GeV}^{-8}$. One might naively expect another constraint coming from the requirement that $M_\sigma$ is a real positive but the critical value for that to be false lies outside the stability conditions: $g^{(0)}_1=20933\mathrm{GeV}^{-8}$.}

This wide spectrum of allowed $g_1(G)$ encompasses different scenarios ranging from the low $g_1$(high $G$) case where, similarly to what happens in the NJLH model, the NJL term is the responsible for D$\chi$SB, to the high $g_1$ (low $G$) scenario the NJL term is too weak to induce it and the physical vacuum is in fact induced by the 6-quark interactions.

These different scenarios can be disentagled by considering the non-vacuum properties of the model. This can be seen in study of the effects of the inclusion of a constant magnetic field \cite{Osipov:2007je,Hiller:2008eh} or of finite temperature \cite{Osipov:2006ev,Osipov:2007mk} and/or chemical potential which are to a large extent dictated by the relative strength of the 4-quark and 8-quark interactions.
Our focus will be on the latter since we are concerned with the phase diagram of QCD.

For $N_f>2$ the bosonization of this effective quark Lagrangian can be done in the path integral formalism with the aid of the introduction of two sets of flavour nonet bosonic fields: $\{\sigma_a,\phi_a\}$, related to the physical scalar and pseudo-scalar mesons,  and $\{s_a,p_a\}$ which are auxiliary fields \cite{Reinhardt:1988xu}. The vacuum-to-vacuum amplitude can then be split in two parts, enabling the separate evaluation of the functional integral over the fermionic degress of freedom and over the auxiliary variables. The former can be done using a symmetry preserving heat kernel scheme (generalized to deal with non-degenerate quark current masses \cite{Osipov:2001nx, Osipov:2000rg, Osipov:2001bj}) whereas the latter can be done using the stationary phase approximation (for the present results it was done to leading order).

The minimization of the effective (or thermodynamic) potential then corresponds to the self consistent solution of two sets of equations (for details see for instance \cite{Osipov:2006ns} and references therein). One of these comes from demanding vanishing vacuum expectation value for the meson fields, the gap equations,
\begin{equation}
\label{StPhaseEqs}
\left\{
\begin{array}[c]{c}%
\Delta_u +G h_u+ \frac{\kappa}{16} h_d h_s +
\frac{g_1}{4}h_u\left(h_u^2+h_d^2+h_s^2\right)+\frac{g_2}{2}h^3_u=0\\
\Delta_d +G h_d+ \frac{\kappa}{16} h_u h_s +
\frac{g_1}{4}h_d\left(h_u^2+h_d^2+h_s^2\right)+\frac{g_2}{2}h^3_d=0\\
\Delta_s +G h_s+ \frac{\kappa}{16} h_u h_d +
\frac{g_1}{4}h_s\left(h_u^2+h_d^2+h_s^2\right)+\frac{g_2}{2}h^3_s=0
\end{array}
\right.,
\end{equation}
while the other comes from the Stationary Phase requirements to integrate out auxiliary fields $\left\{s_a,p_a\right\}$,
\begin{align}
\label{GapEqs}
\left\{
\begin{array}{r}
\frac{1}{2}h_u+\frac{N_c M_u}{24\pi^2}
\left( 6 I_0- \left(\Delta_{us}+\Delta_{ds}+3\Delta_{ud}\right)I_1 \right)=0\\
\frac{1}{2}h_d+\frac{N_c M_d}{24\pi^2}
\left( 6 I_0- \left(\Delta_{us}+\Delta_{ds}-3\Delta_{ud}\right)I_1 \right)=0\\
\frac{1}{2}h_s+\frac{N_c M_s}{24\pi^2}
\left( 6 I_0+ 2\left(\Delta_{us}+\Delta_{ds}\right)I_1\right)=0
\end{array}
\right. .
\end{align}
Here $\Delta_i=M_i-m_i$, with $M_i$, the dynamical mass of the quark species $i$ and the $\Delta_{ij}$ are differences of squared masses $\Delta_{ij}=M^2_i-M^2_j$.
The $I_i$ correspond to the averaged sum of $I_i=1/3\sum_{i=u,d,s}J_i$, where the $J_i$ are quark 1-loop euclidean momentum integrals with $i+1$ vertices. For their regularization we chose to use two Pauli-Villars subtractions resulting in the definition:
\begin{align}
J_i&=16 \pi ^2 \Gamma(i+1)\int \frac{\mathrm{d}^4 p_E}{(2\pi )^4}\hat{\rho}_\Lambda\frac{1}{\left(p_E^2+M^2\right)^{i+1}}\nonumber\\
\hat{\rho}_\Lambda &=1-(1-\Lambda^2\frac{\partial}{\partial M^2})e^{\Lambda^2\frac{\partial}{\partial M^2}}
\end{align}
Integrals of different orders can be easily related through recursion relations: $J_{i+1}=-\partial_{M^2} J_i$.
Their generalization to take into account the effect of finite temperature and/or chemical potential is done through the introduction of the Matsubara frequencies as per usual.

The thermodynamic (or effective) potential can be obtained by integration of the gap equations, resulting in (see \cite{Hiller:2008nu} for details):
\begin{align}
\label{Veff}
\Omega=&\frac{1}{32}\left(8 G\left(h_u^2+h_d^2+h_s^2\right)+2 \kappa h_u h_d h_s +3 g_1 \left(h_u^2+h_d^2+h_s^2\right) + 6\left(h_u^4+h_d^4+h_s^4\right) \right)\nonumber\\
& +\frac{3 N_c}{8 \pi^2} \left(J_{-1}(M_u,\Lambda,\mu_u,T)+J_{-1}(M_d,\Lambda,\mu_d,T)+J_{-1}(M_s,\Lambda,\mu_s,T)\right).
\end{align}
The quark loop integral contribution, the $J_{-1}$ integrals are given by:
\begin{align}
J_{-1}(M,\Lambda,T,\mu)&=-8\int^\infty_0\mathrm{d}|\overrightarrow{p_E}|\left(|\overrightarrow{p_E}|^2\hat{\rho}_\Lambda E_p\right)-\frac{8}{3}\int^\infty_0\mathrm{d}|\overrightarrow{p_E}|\left(|\overrightarrow{p_E}|^4\hat{\rho}_\Lambda\frac{n_q+n_{\overline{q}}}{E_p}\right) -\left.\left[\ldots\right]\right|_{M=0}\nonumber\\
&=J_{-1}(M,\Lambda)-\frac{8}{3}\int^\infty_0\mathrm{d}|\overrightarrow{p_E}|\left(|\overrightarrow{p_E}|^4\hat{\rho}_\Lambda\frac{n_q+n_{\overline{q}}}{E_p}\right),
\end{align}
where the last subtraction corresponds to the integrals evaluated with $M=0$.
In the last line we have made explicit the separation between the vacuum and the medium contribution. The latter involves (anti-)quark occupation numbers which are defined as:
\begin{align}
n_q=\frac{1}{1+e^{\frac{E_p-\mu}{T}}},\quad
n_{\overline{q}}=\frac{1}{1+e^{\frac{E_p+\mu}{T}}}.
\end{align}

\section{Results}

As was mentioned above there is a wide range of allowed values for $g_1$ (and $G$, as they have to be changed in consonance for the meson spectra to be left more or less unchanged) if we take into account only vacuum stability restrictions. In \cite{Osipov:2006ev,Osipov:2007mk} it was shown that the temperature behaviour at vanishing chemical potential changes dramatically over this range. The OZI violating eight quark interactions dictate the temperature, slope and nature of the chiral transition: higher $g_1$ results in lower $T_c$, with a steeper slope and above a certain critical value one obtains a first order transition (which is made possible by the above described scenario of a physical vacuum induced by the 't Hooft term in a potential that would otherwise correspond to a distorted Wigner-Weyl phase). As there is growing evidence coming from lattice calculations \cite{Aoki:2006we, Aoki:2006br} that the transition at $\mu=0$ is crossover this can be interpreted as another upper bound for $g_1$ \footnote{Using the already mentioned fittind procedure we obtain $g^{(3)}_{1}=8703\mathrm{GeV}^{-8}$ which is lower than the upper bound required by vacuum stability.}.

\begin{figure}[htp]
\label{mudep}
\begin{tabular}{cc}
\includegraphics[width= 0.25 \columnwidth]{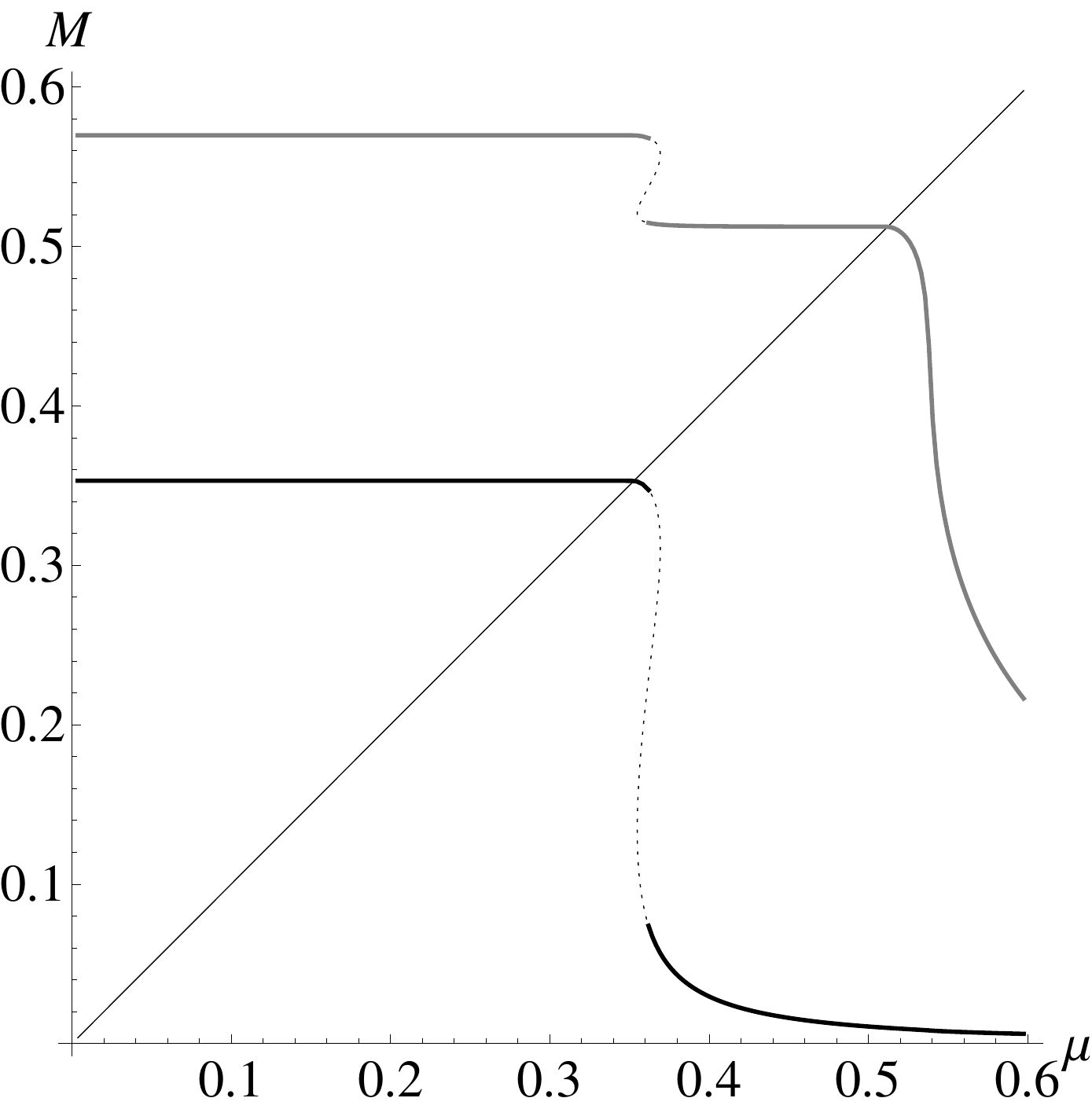}
&
\includegraphics[width= 0.25 \columnwidth]{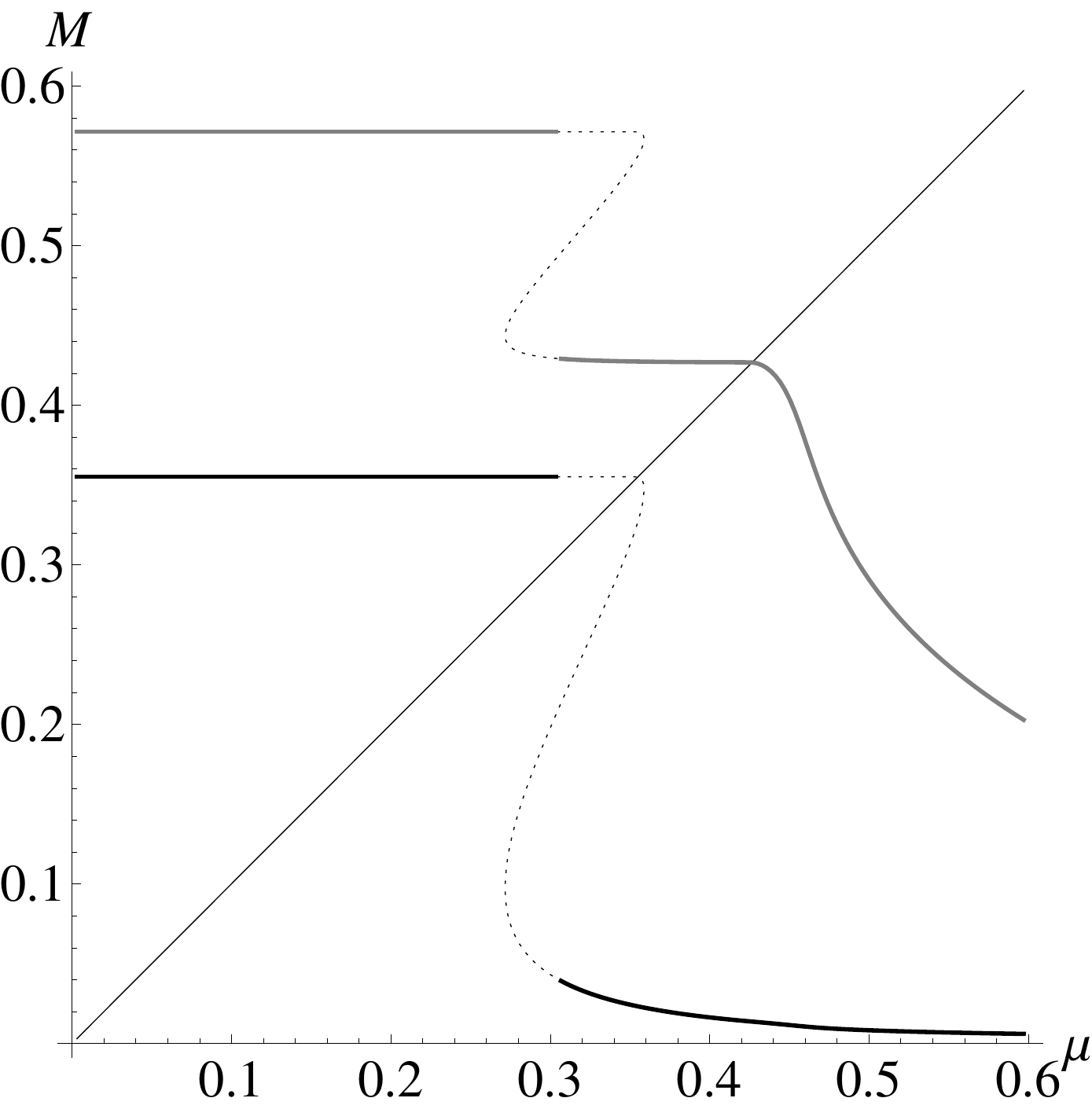}
\end{tabular}
\caption{Dynamical mass of the quarks as a function of chemical potential at zero temperature ($\left[M\right]=\left[\mu\right]=\mathrm{GeV}$). Left panel refers to the weak coupling case set a) of Table \ref{ParamSets} whereas the right panel refers to set b). The upper grey thick lines refer to $M_s$, the black thick lines refer to $M_u$ and the dotted line refers to unstable solutions which are skipped in the process of the first order transition. The diagonal line refers to $M=\mu$.}
\end{figure}

In the case for zero temperature and finite chemical potential (considered equal for all quark species $\mu=\mu_u=\mu_d=\mu_s$), on the other hand we always find a first order chiral transition for some critical value $\mu_c$ which is lower for higher $g_1$. The analysis of the dependence of the quarks dynamical mass as a function of the chemical potential for different values of $g_1$ (see Fig. \ref{mudep}) reveals that, below a certain critical value $g^{(4)}_{1}$, the model describes a low density phase of massive quarks (notice that on the left panel the line refering to $M_{u,d}(\mu)$ crosses the diagonal before the first order transition jump). As this is deemed unphysical (see for instance \cite{Buballa:2003qv}) this sets a lower bound for the interaction strength \footnote{For the chosen fits  we have $g^{(4)}_{1}=2423\mathrm{GeV}^{-8}$ which is also inside the range allowed by vacuum stability}.

\begin{figure}[htp]
\label{Mdyndens}
\begin{tabular}{cc}
\includegraphics[width= 0.25 \columnwidth]{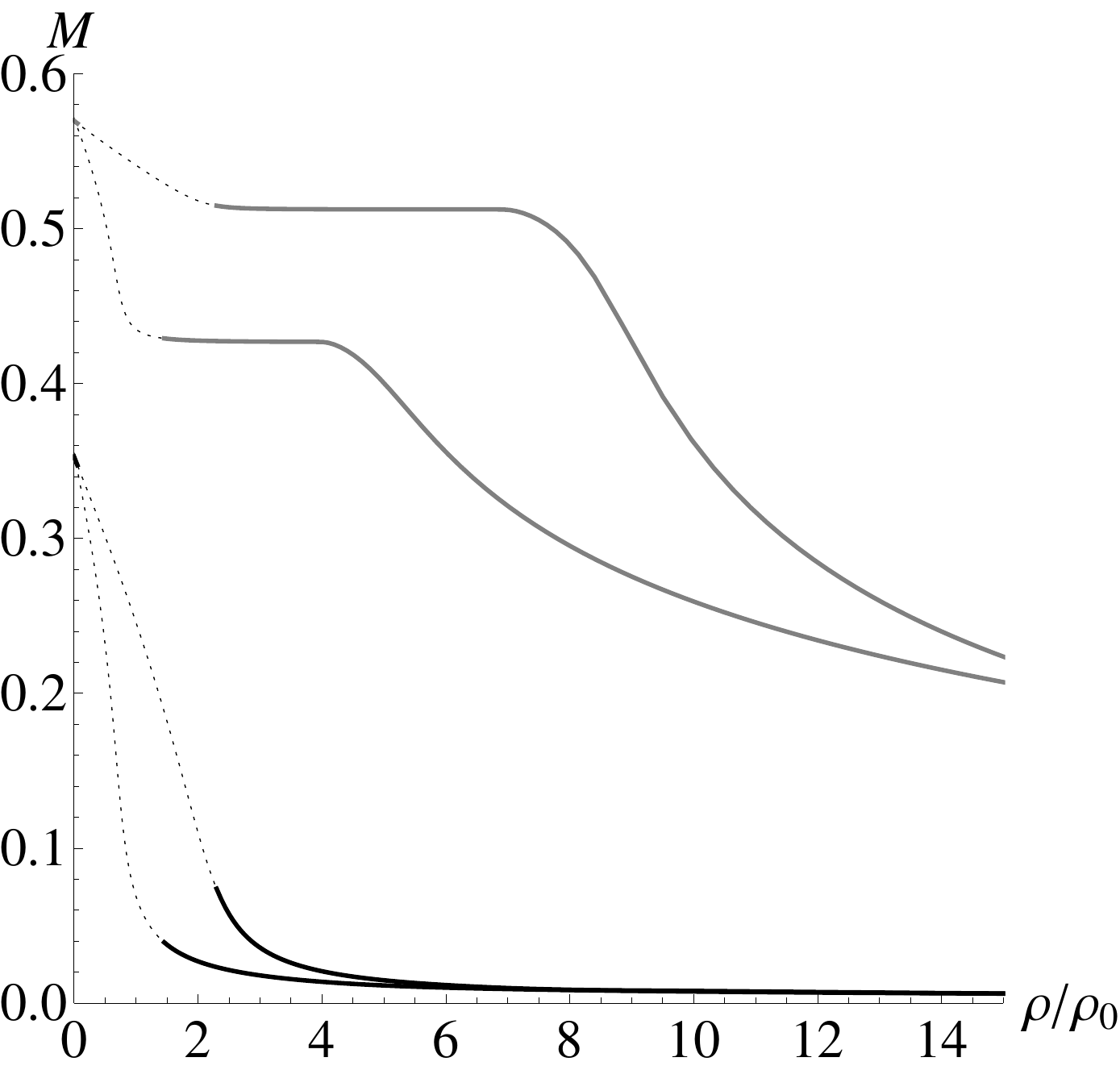}
&
\includegraphics[width= 0.25 \columnwidth]{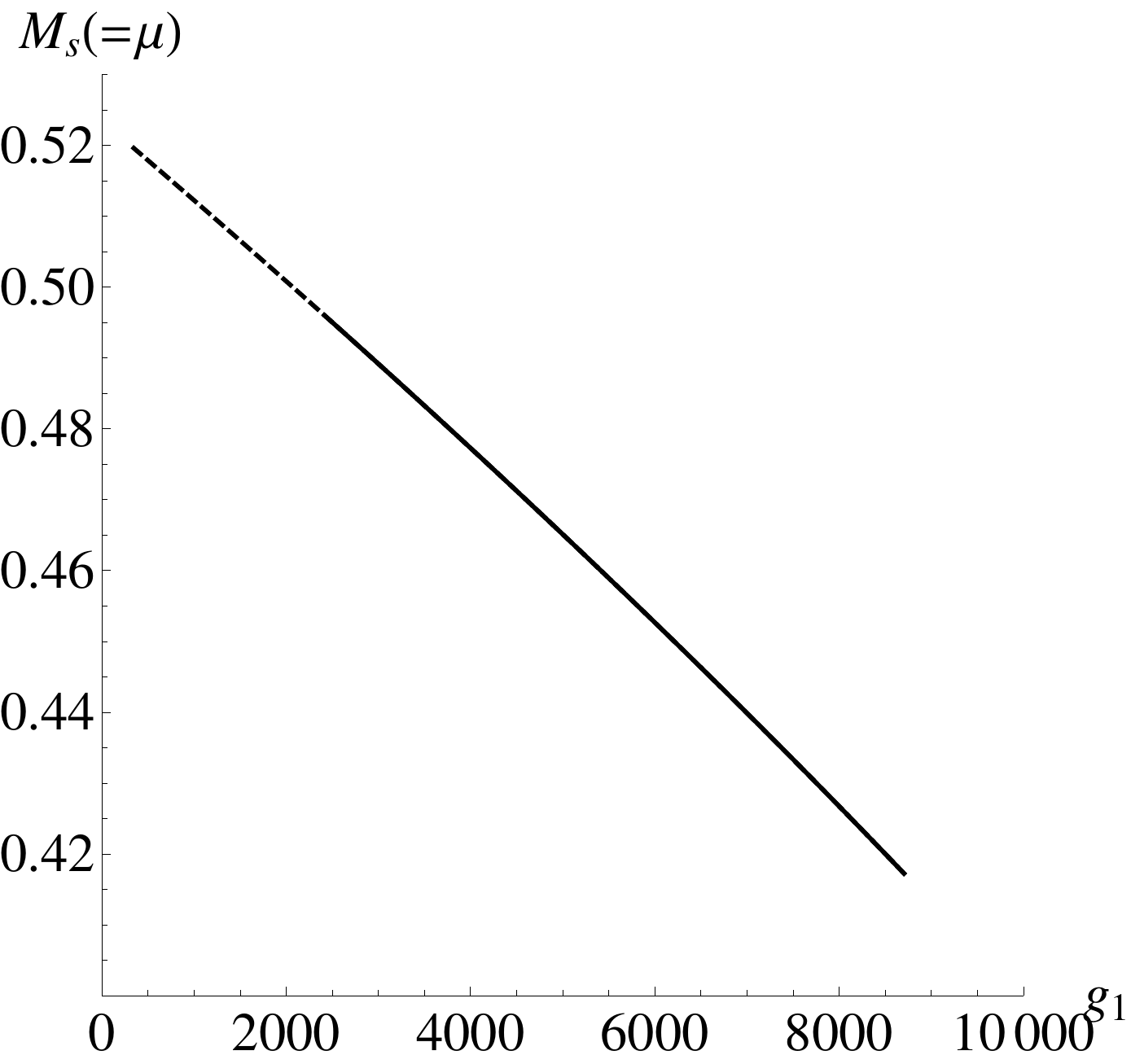}\\
a) & b)
\end{tabular}
\caption{\textbf{a)} Dynamical mass of the quarks ($\left[M\right]=\mathrm{GeV}$) as a function of baryonic density (divided by $\rho_0= 0.17 \mathrm{fm}^{-3}$) at zero temperature. The upper grey and black lines correspond respectively  to $M_s$ and $M_{u,d}$ in the weak coupling parameter set (the lower correspond to stronger coupling). The dotted lines correspond to unstable solutions which are skipped by the first order transition. \textbf{b)} Chemical potential above which the density of strange quarks becomes non-vanishing as a function of $g_1$. The dashed line corresponds to the range of $g_1$ values alllowed by vacuum stability but excluded by the requirement of the unexistence of a low-density phase of massive quarks. The full line goes through the allowed values up to the limit that corresponds to the requirement of crossover transition at zero chemical potential.}
\end{figure}

Ploting the dynamical mass as a function on the bayonic density (as can be seen in the left panel of Fig. \ref{Mdyndens}) we see furthermore that the jump in $M_s$ is higher and occurs at lower densities for stronger coupling. Furthermore the critical chemical potential for non-vanishing strange quark density is lower for higher $g_1$. Both these results may have interesting implications in the physics of compact stars.

\begin{figure}[htb]
\label{PhaseDiagram}
  \includegraphics[width= 0.3 \columnwidth]{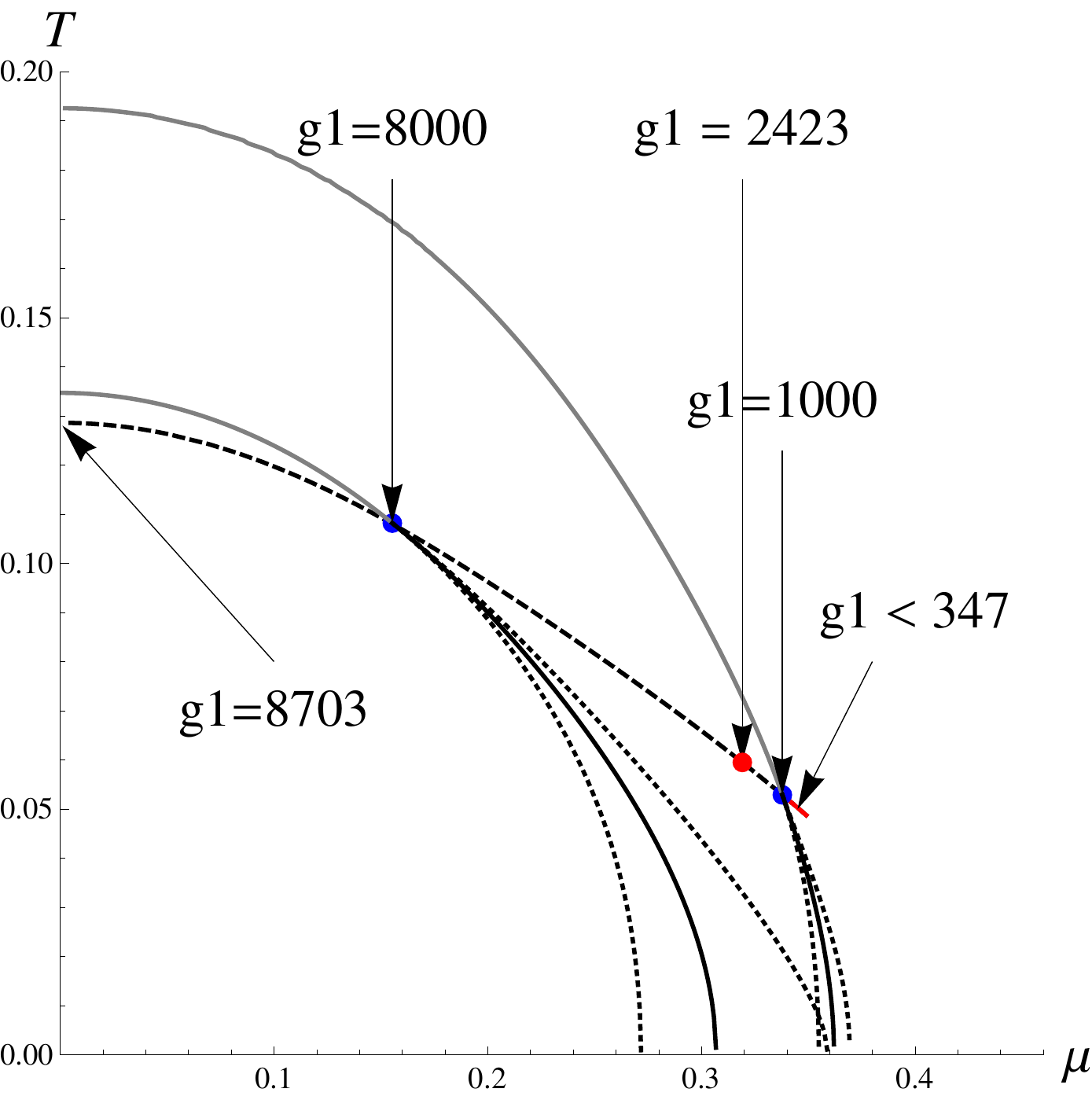}
  \caption{Phase diagram for diferent values of $g_1$, given in units of $\mathrm{GeV}^{-8}$ . Temperature, $T$, and chemical potential, $\mu$, are given in $\mathrm{GeV}$. The gray lines correspond the crossover transition for the two parameter sets from Table \ref{ParamSets} (outermost corresponds to weaker coupling). The blue dots, the full and dotted black lines correspond (again for the two sample parameter sets) respectively to: CEP, first order transition and spinodals. The red dot marks the position of the CEP for the critical value below which a low density gas of massive quarks is obtained $T=0$, $g_{1}=2423\mathrm{GeV}^{-8}$. The dashed line corresponds to the range of allowd CEP.}
\end{figure}

For finite $T$ and $\mu$ (see Fig. \ref{PhaseDiagram}) we find that the crossover/first order chiral transition line and the position of the Critical End Point (CEP) are highly sensitive to the chosen value for $g_1$. As the transition as a function of $T$ is steeper for higher $g_1$ the critical $\mu$ for it to become fist order is lower for higher $g_1$: the CEP moves to the left in the phase diagram with increasing OZI-violating eight quark interactions. Furthermore in the studied cases the transition line, $T_c(\mu)$, for higher $g_1$ lies completely inside that of lower $g_1$.

\begin{table}[htb]
\begin{tabular}[c]{l||r|r|r|r|r|r|r|r|r }
\hline
     \tablehead{1}{r}{b}{Sets}
   & \tablehead{1}{r}{b}{$m_u$\\($\mathrm{MeV}$)}
   & \tablehead{1}{r}{b}{$m_s$\\($\mathrm{MeV}$)}
   & \tablehead{1}{r}{b}{$M_u$\\($\mathrm{MeV}$)}
   & \tablehead{1}{r}{b}{$M_s$\\($\mathrm{MeV}$)}
   & \tablehead{1}{r}{b}{$\Lambda$\\($\mathrm{MeV}$)}
   & \tablehead{1}{r}{b}{$G$\\($\mathrm{GeV}^{-2}$)}
   & \tablehead{1}{r}{b}{$\kappa$\\($\mathrm{GeV}^{-5}$)}
   & \tablehead{1}{r}{b}{$g_1$\\($\mathrm{GeV}^{-8}$)}
   & \tablehead{1}{r}{b}{$g_2$\\($\mathrm{GeV}^{-8}$)}\\
\hline\hline
a & $5.9$ & $186$ & $359$ & $554$ & $851$ & $10.92$ & $-1001$ & $1000^\ast$ & $-47$\\
\hline
b & $5.9$ & $186$ & $359$ & $554$ & $851$ & $7.03$ & $-1001$ & $8000^\ast$ & $-47$\\
\hline
\end{tabular}
\caption{Parameter sets obtained by fits to the $\pi$ , $\kappa$, $\eta^\prime$ and $a_0$ masses, as well as the $\pi$ and $K$ weak decay constants. The value for OZI violating part of the eight quark interactions (marked with $^\ast$) is not fitted.}
\label{ParamSets}
\end{table}

\section{Conclusions}

The eight quark interactions necessary to stabilize the ground state in three flavor NJLH model, dictate the phase diagram of the model. Their strength can be constrained using considerations on vacuum stability and the requirements of crossover at vanishing chemical potential as well as the unexistence of a low density phase of massive quarks.


\begin{theacknowledgments}

This work has been supported in part by grants of Fundação para a Ciência e Tecnologia, FEDER, OE, POCI 2010,
CERN/FP/83510/2008, SFRH/BD/13528/2003 and Centro de Física Computacional, unit 405.
We acknowledge the support of the European Community-Research Infrastructure Integrating Activity Study of
Strongly Interacting Matter (acronym HadronPhysics2, Grant Agreement n.  227431) under the Seventh Framework
Programme of EU.

\end{theacknowledgments}



\bibliographystyle{aipproc}   

\bibliography{Hadron2009Proceedings3}

\IfFileExists{\jobname.bbl}{}
 {\typeout{}
  \typeout{******************************************}
  \typeout{** Please run "bibtex \jobname" to optain}
  \typeout{** the bibliography and then re-run LaTeX}
  \typeout{** twice to fix the references!}
  \typeout{******************************************}
  \typeout{}
 }

\end{document}